\documentclass[aps,pra,twocolumn,superscriptaddress]{revtex4-2}\usepackage[utf8]{inputenc}
\usepackage{amsmath}

\usepackage{float}
\makeatletter
\let\newfloat\newfloat@ltx
\makeatother

\usepackage{bm}
\usepackage{amssymb,amsfonts,latexsym,fancyhdr,graphicx,epstopdf,algorithm,algorithmic}

\usepackage[utf8]{inputenc}
\usepackage{soul}

\usepackage[english]{babel}
\usepackage{graphicx}
\usepackage {subfigure}
\usepackage[braket, qm]{qcircuit}
\usepackage{xcolor}

\usepackage[colorlinks=true,
            linkcolor=blue,
            citecolor=red,
            urlcolor=magenta]{hyperref}

\usepackage{verbatim}

\DeclareMathOperator{\Tr}{Tr}

\begin{document}

\title{Bayesian mitigation of measurement errors in multiqubit experiments}

    \author{F. Cosco}
    \email{francesco.cosco@vtt.fi}
    \affiliation{Quantum algorithms and software, VTT Technical Research Centre of Finland Ltd, Tietotie 3, 02150 Espoo, Finland}
    \author{F. Plastina}
    \affiliation{Dipartimento di Fisica, Universit\`a della Calabria, 87036 Arcavacata di Rende (CS), Italy}
    \affiliation{INFN, Gruppo Collegato di Cosenza, 87036 Arcavacata di Rende (CS), Italy}
    \author{N. Lo Gullo}
    \email{nicolino.logullo@unical.it}
    \affiliation{Dipartimento di Fisica, Universit\`a della Calabria, 87036 Arcavacata di Rende (CS), Italy}
    \affiliation{INFN, Gruppo Collegato di Cosenza, 87036 Arcavacata di Rende (CS), Italy}
    \affiliation{4ICAR-CNR, Consiglio Nazionale delle Ricerche, Rende, Cosenza, Italy}

\begin{abstract}
 {In \href{https://doi.org/10.1103/PhysRevA.108.L060402}{Phys. Rev. A {\bf 108}, L060402 (2023)}, we introduced a Bayesian measurement error mitigation algorithm, which leverages complete information from the readout signal, and validated the protocol on a quantum device with five superconducting qubits. Here, we present an improved algorithm's implementation, tailored for multiqubit experiments on near-term superconducting qubit quantum devices. 
In particular, we provide a detailed algorithm workflow, from calibrating the detector response functions to the post-processing of measurement outcomes, offering a computationally efficient solution for the output size typical of current quantum computing devices.  
We show how the numerical representation of the noise function affects the performance of the error mitigation algorithm and test the convergence criteria. We benchmark our protocol on actual quantum computers with superconducting qubits, where the readout signal encodes the measurement information as unprocessed analog data before qubit state assignment. Finally, we compare the performance of our algorithm against other measurement error mitigation methods, such as iterative Bayesian unfolding and the Mthree method, and show how our method can be integrated on top of other readout error mitigation protocols. }
\end{abstract}

\maketitle

\section{Introduction}
Despite rapid advancements in quantum computing technologies, practical applications of noisy intermediate-scale quantum (NISQ) processing units (QPUs) are still out of reach. While QPU manufacturers are constantly improving quality metrics \cite{Hyyppa2022,Marxer2023,reglade2024}, be it decoherence times, gate and readout fidelities, a lot of effort is being invested in developing software tools to shrink the gap between acceptable noise levels and utility-scale algorithms \cite{mundada2023,skoric2023}. Quantum error mitigation (QEM) embodies this goal, although a comprehensive and universally accepted definition for the field is lacking \cite{Cai2023}. In general, QEM encompasses all those algorithmic schemes that reduce noise and errors in the expectation value of some figure of merit, such as the average of observables or quantum state probabilities. This is typically achieved by post-processing the output from an ensemble of circuit runs, either the same circuit or a family of properly designed ones \cite{filippov2024scalability}. This statistical approach starkly contrasts with quantum error correction, which needs to detect errors through single shot measurements. Among the most advanced techniques, zero-noise extrapolation \cite{Li2017} with probabilistic error amplification, probabilistic error cancellation \cite{Temme2017}, and tensor network error mitigation \cite{filippov2023scalable} have recently been found to be successful in the largest available QPUs \cite{kim2023evidence,van2023probabilistic,bluvstein2024logical, fischer2024dynamical}.
Some specialized techniques aim to target specific sources of errors, such as measurement error mitigation strategies. These strategies are designed to improve the readout step of quantum algorithms execution by using a model of the measurement error, which is a result of the faulty measurement process.


\begin{figure}[!t]
\begin {center}
   \includegraphics[width=\columnwidth]{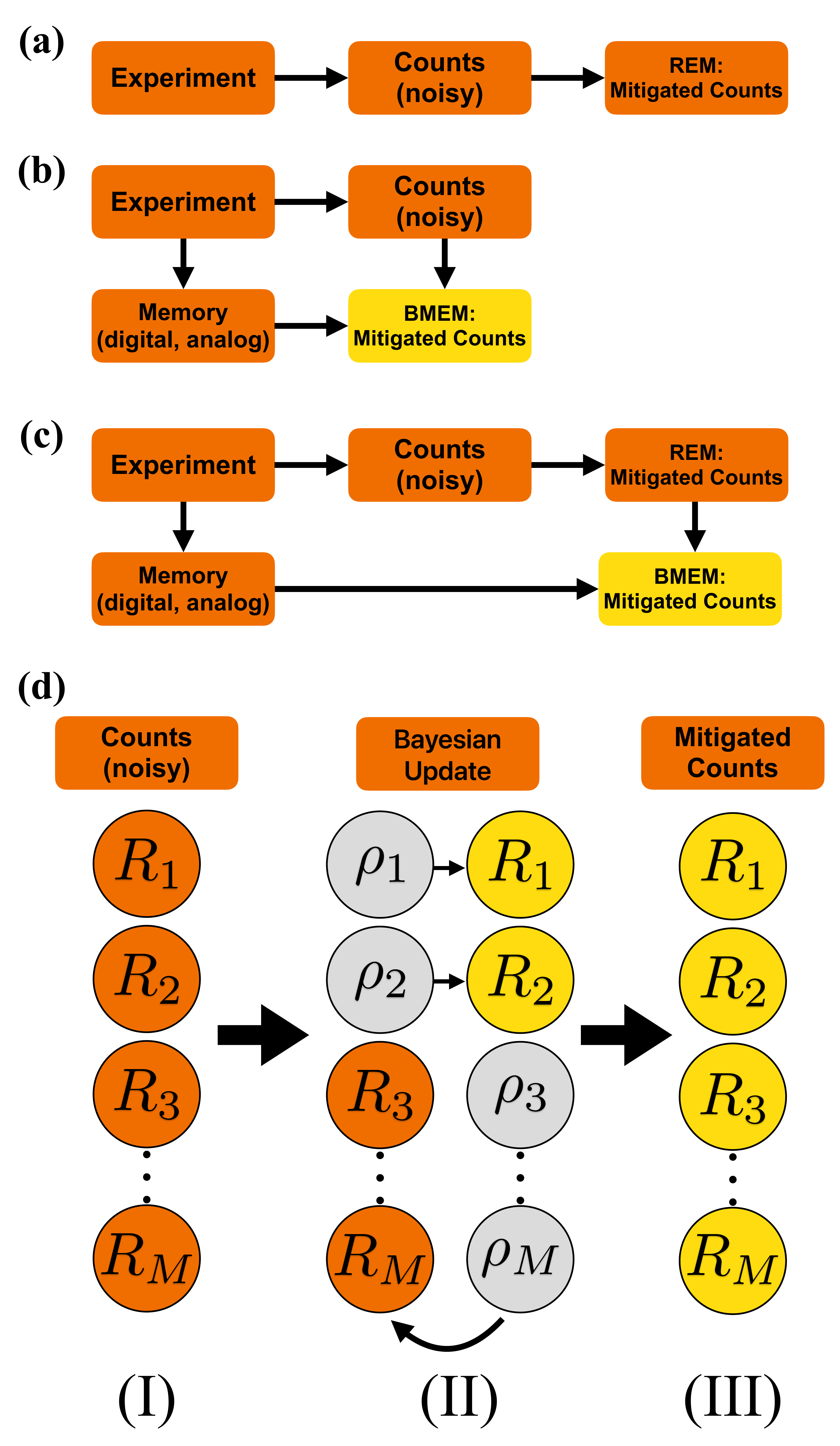}
\end{center}
\caption {{(a) Workflow of standard readout error mitigation. (b) Workflow of Bayesian measurement error mitigation. (c) Workflow of Bayesian measurement error mitigation when integrating other readout error mitigation techniques.}
(d) Pictorial sketch of the heuristic algorithm to perform Bayesian measurement error mitigation in a multiqubit experiment. In the first step (I), we collect the noisy bitstrings counts and reduce the \(2^{\mathrm{N_q}}\) outcome subspace to mitigate to the \(M\) measured noisy counts for which we have at least one measurement. In step (II), we select two populations to use as variables while keeping all others constant, and apply Bayesian measurement error mitigation to obtain two new mitigated values. The Bayesian update is performed for all possible pairs. Step (II) is repeated until a specified convergence criterion or exit condition is met, resulting in the full set of mitigated values (III).}
\label {fig:sketch}
\end{figure}

Most readout mitigation techniques are based on some form of detector or measurement tomography \cite{feito2009,Lundeen2009,Maciejewski2020,Cattaneo2023} to obtain the assignment, noise, or confusion matrix. The readout error is then addressed in various ways. One straightforward strategy involves manipulating the outcome count statistics by multiplying the inverse of the confusion matrix by the measured probability vector of the quantum state \cite{Bravyi2021}. However, this approach can produce nonphysical values, such as negative probabilities, an issue that can be taken into account by incorporating a convex optimizer into the mitigation pipeline \cite{ChenPRA2019}. Alternatively, some proposals have developed machine learning-based techniques to perform the assignment process \cite{Navarathna2021,Lienhard2022,Azad2022,ahmad2024mitigating}.

In a recent work \cite{Cosco2023}, we have introduced an efficient and accurate readout measurement scheme for single- and multi-qubit states based on Bayesian inference, inspired by successful application of Bayesian inference in quantum metrology and sensing \cite{Dinani2019,Puebla2021} or parameter estimation for quantum circuits and states~\cite{Leibfried2005,Teklu2009,Lundeen2009,Cassemiro_2010,Paesani2017,Laverick2021,Smerzi2021,Duffield2022}. 
{By bypassing the need to invert the noise matrix, our method leverages knowledge of the detector's faulty response and employs Bayesian inference to estimate probability distributions for each qubit state.}
Moreover, such an approach is not limited to using binary outcomes from the detector, but it can exploit the detector unprocessed analog readout data before any binary assignment is made.  The idea of using analog data has also found traction in quantum error correction, where it has been shown to lead to improved decoders and logical qubit error rates \cite{pattison2021improved,ali2024reducing}. Still, all measurement error mitigation schemes suffer some form of scalability issue when
moving to many-qubit systems, as intuitively the noise matrix size scales exponentially with the number of qubits. Often, this issue is addressed by not using the full assignment matrix, or its inverse, and instead working in a subspace defined by the noisy outcomes \cite{nation2021, Yang2022}.
In this work, we aim to tackle thoroughly the scaling problem when aiming to mitigate measurement errors with our formerly introduced Bayesian readout and further show how the use of the readout analog data leads to increased performances. Beyond showing a solid proof of principle of the method, we make a full scaling analysis of the algorithm and provide an estimate of the classical computation time needed to post-process the measurement results obtained from currently available superconducting QPUs.

{The paper is organized as follows. In Section \ref{theory}, we present the theoretical foundations necessary to model measurement errors in quantum computers. We demonstrate how a Bayesian approach can be employed to infer the true outcomes from noisy measurements, assuming prior knowledge of the detector noise matrix. Additionally, we introduce a novel algorithm for implementing Bayesian measurement error mitigation in the context of multi-qubit measurements. The proposed method includes a detailed, step-by-step procedure suitable for current quantum hardware. We also describe how to adapt this framework to various types of measurement data, including binary outcomes (i.e., "0"s and "1"s) and unprocessed analog signals, such as IQ readout data for superconducting QPUs \cite{krantz2019quantum}.}

In Section \ref{scaling}, we benchmark our implementation of Bayesian measurement error mitigation by running quantum algorithms on actual quantum hardware. We demonstrate how the algorithm can be efficiently applied to improve readout accuracy in multi-qubit experiments and show how the performance of the mitigation scheme scales with both the number of qubits and the number of measurements for the chosen examples.

  { Section \ref{analog} focuses on applying the Bayesian measurement error mitigation scheme to analog detector data. We show that utilizing the full IQ-readout signal—rather than the binary outcomes—can significantly improve the performance of the mitigation algorithm in multi-qubit scenarios. Additionally, we discuss the  computational cost and convergence criteria of our implementation of the algorithm for multi-qubit Bayesian measurement error mitigation and provide, as an example, the computational times required for different state preparations. In Section \ref{comparison}, we compare our Bayesian methods against two other measurement error mitigation schemes and show how it could be integrated with them. Throughout this work, we show the results obtained by executing the quantum circuits on three superconducting qubit devices: VTTQ20, VTTQ50, and IBM127 (Nazca).
Finally, we present the conclusions and outline some possible future research directions and applications in Section \ref{end}.}  

\section{Theoretical background}
\label{theory}

Let's start with a mathematical description of a noisy measurement process for a single qubit. Let $s \in \{0, 1\}$  be the measurement outcome observed in a single experiment. In the absence of measurement errors, the probability to register either one is fully determined by the quantum state of the system. Formally, a conventional single qubit measurement is modeled with a Positive-Operator Valued Measure (POVM) with two effects $\{M_0 =|0 \rangle \langle 0|, M_1 = |1 \rangle \langle 1|\}$. Consequently, the probability of observing a measurement outcome $s$, when the quantum system is in a state $\hat \rho$, can be calculated as $\mathrm{P}(s) =\Tr (\hat \rho M_s ) \equiv \rho_s$. 

However, in a real setup, the readout step is noisy and errors emerge in the process of collecting the signal from the detector and in the assignment of the correct qubit state. In practice, we can model the experimental noisy measurement through a set of noisy effects $M^{\mathrm {noisy}}_i = \sum_j \langle i | \Lambda |j \rangle M_j = \sum_j \Lambda_{ij} M_j$, where $\Lambda$ is the noise matrix and  $\Lambda_{ij}$ is the probability of observing/assigning outcome $i$ when the true outcome is $j$.
Therefore, after repeated measurements, we have direct access to the noisy probabilities
\begin{equation}
\rho^{\mathrm {noisy}}_s = \Lambda_{sj} \rho_j,
\label{cond-prob-noise}
\end{equation}
which are connected to the true qubit state probabilities through the noise matrix  $\Lambda$. Thus, the goal of measurement error mitigation is to reconstruct the true qubit state probabilities after measuring $\rho^{\mathrm {noisy}}_i$, as in the workflow of Fig. \ref{fig:sketch} (a). Assuming that we have access to the noise matrix $\Lambda$, from Eq. \eqref{cond-prob-noise}, it is straightforward to attempt to invert the noise matrix and use it to reconstruct the correct qubit state probability vector as $\rho = \Lambda^{-1} \rho^{\mathrm {noise}}$. 
However, this simple measurement error mitigation strategy might result in a probability vector containing negative values due to statistical errors. This issue is often accounted for by employing a convex optimizer which project the corrected vector to a physical one \cite{Maciejewski2020,ChenPRA2019}.

To avoid these issues, we propose here a strategy to bypass the need to invert the noise matrix by noting how the measured outcome $\rho^{\mathrm {noisy}}_s$, as introduced in Eq. \eqref{cond-prob-noise}, can be read as the conditional probability of measuring $s$ when the correct qubit state probabilities are $\rho_0$ and $\rho_1$, i.e. $\rho^{\mathrm {noisy}}_s = \mathrm{P} (s|\rho_0, \rho_1)$.

Thus, we can reconstruct a probability distribution for the correct probabilities $\rho_0$ and $\rho_1$  through Bayes' theorem as $\mathrm{P}(\rho_0, \rho_1|s)\propto  \left [\Lambda_{s0} \rho_0  +\Lambda_{s1}\rho_1 \right ]\mathrm{P}(\rho_0, \rho_1)$. It is convenient to follow an iterative approach and consider individually the $\mathrm N_{\mathrm{shots}}$ values collected after each measurement. In this way, we can update the conditional probability for the quantum state probabilities $\rho_0$ and $\rho_1$ shot after shot as
\begin{equation}
\mathrm{P}(\rho_0,\rho_1|s^{(n)})\propto  \left [\Lambda_{s0} \rho_0  +\Lambda_{s1}\rho_1 \right ]\mathrm{P}(\rho_0, \rho_1| s^{(n-1)}),
\label{bayes-update}
\end{equation}
where $s^{(n)} \in \{0, 1\}$ is the bit registered in the n-th experiment, and $\mathrm{P}(\rho_0, \rho_1 | s^{(n-1)})$ is the prior probability distribution which corresponds to the posterior distribution calculated at the previous step. The idea is to update the prior probability distribution after each iteration step with the newly calculated posterior probability distribution. When assuming no prior knowledge about the system state, we start the iteration with a uniform prior probability distribution $\mathrm { Pr}(\rho_0,\rho_1| s^{(0)}) = \mathrm{const}$.

The end result of the Bayesian measurement error mitigation scheme
is the probability distribution $\mathrm{P}(\rho_0, \rho_1| s^{(\mathrm N_{\mathrm {shots}})})$ which represents the posterior probability distribution for the qubit state probabilities coherent with all the collected measurement data. 
 {Finally, this distribution can be used to estimate the ideal qubit state probabilities $\mathrm{P}{(0)}$ and $\mathrm{P}{(1)}$. The workflow of the mitigation algorithms follows Fig. \ref{fig:sketch} (b).}

In our previous work ~\cite{Cosco2023}, we proposed to use the first moment of the distribution $\mathrm{P}(\rho_0,\rho_1| s^{(\mathrm N_{\mathrm {shots}})})$, namely $\mathrm{P}{(0)}=\int d\rho_0 d\rho_1\;\rho_{0}\mathrm{P}(\rho_0,\rho_1| s^{(\mathrm N_{\mathrm {shots}})})$ and similarly for $\mathrm{P}{(1)}$. Nevertheless, we found that results are highly improved by choosing the values $\rho_{0/1}$ for which the probability $\mathrm{P}(\rho_0,\rho_1| s^{(\mathrm N_{\mathrm {shots}})})$ has its maximum , i.e. $(\mathrm{P}{(0)},\mathrm{P}{(1)})=\underset{\rho_0, \rho_1}{\arg\max} \, \mathrm{P}(\rho_0,\rho_1| s^{(\mathrm N_{\mathrm {shots}})})$. 

With these ingredients, we can extend the Bayesian measurement error mitigation scheme to the multi-qubit case with relative ease. For a system of $\mathrm N_q$ qubits, the measurement outcome obtained from a single experiment is a $\mathrm N_q$ digits string, $\mathbf s \in \{0, 1\}^{\mathrm N_q}$. Equivalent to Eq. \eqref{cond-prob-noise}, the conditional probability of observing the multi-qubit string  $\mathbf s$ is related to the correct outcome through a multi-qubit noise matrix as

\begin{equation}
\mathrm{P} (\mathbf s | \bm \rho \equiv \rho_0,...,\rho_{2^{N_q}-1}) = \Lambda_{\mathbf {sj}} \rho_{\mathbf j} .
\label{cond-prob-mq}
\end{equation}

\begin{figure}[!t]
\begin {center}
\includegraphics[width=0.95\columnwidth]{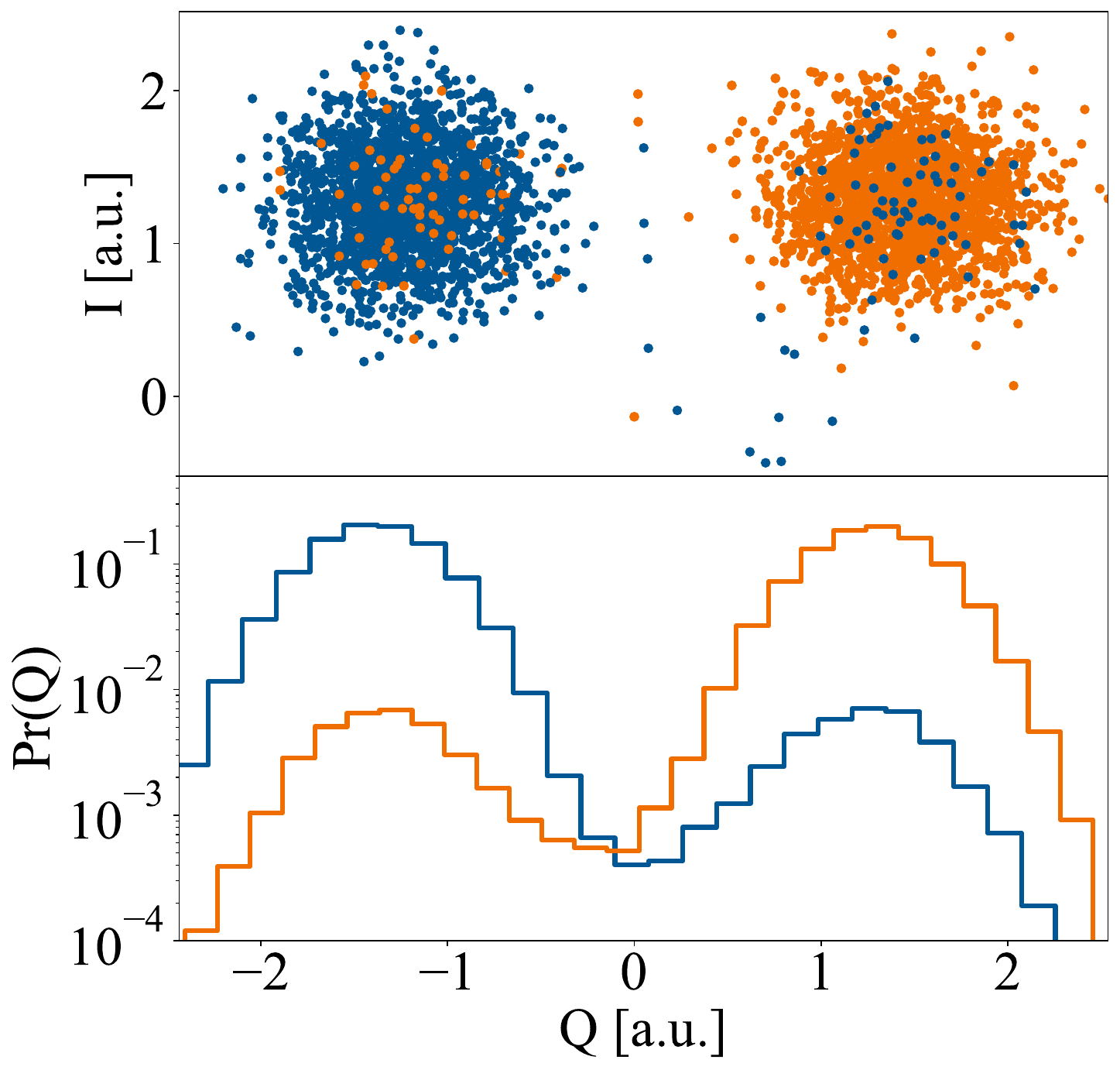}

\end{center}
\caption {(a): (Top) Measurement clouds in the IQ plane, where each point corresponds to a single shot. The blue cloud represents measurements of the "0" state, i.e., when no operation is performed on the qubit besides the measurement. The orange cloud represents measurements of the "1" state, where a single X-gate is applied to the qubit before measurement. (Bottom) Projected histograms from the IQ plane onto the Q-axis.}
\label {fig:clouds}
\end{figure}

From the formal expression of Eq. \eqref{cond-prob-mq}, we can see how the number of terms appearing in the noisy conditional probability increases exponentially with the number of qubits composing the measured string $\mathbf s$. Nonetheless, from a purely mathematical point of view, the Bayesian inference of the correct multi-qubit state probabilities can be achieved in a similar fashion to the single qubit case by employing the iterative Bayesian update in the higher dimension, i.e.
\begin{equation}
\mathrm{P}(\bm \rho | \mathbf s^{(n)}) \propto  \Lambda_{\mathbf s^{(n)} \mathbf {j}} \rho_{\mathbf j} \mathrm{P}(\bm \rho| \mathbf s^{(n-1)}),
\label{eq:many-qubit-bayes}
\end{equation}
 within the parameters space which satisfies the normalisation condition $\sum_{\mathbf i} \rho_{\mathbf i} = 1$. However, since the computational cost increases  exponentially with the number of qubits such a scheme becomes of limited utility beyond a few qubits.

For example, for a system with \({\mathrm{N_q}}\) qubits, Eq.~\eqref{eq:many-qubit-bayes} requires handling a multi-parameter probability distribution \(\mathrm{P}(\bm{\rho}|\mathbf{s}^{(n)})\) at each iteration step of the Bayesian update. This distribution involves $2^{\mathrm{N_q}}$ variables $\rho_{\mathbf i}$, each sampled within the interval $[0,1]$. If each variable is sampled with \(n_p\) points, the total probability distribution would necessitate a mesh-grid with $n_p^{2^{\mathrm{N_q}}}$ points. Even though the normalization condition reduces the effective number of independent coordinates the computational cost remains high and out of reach.

Additional complications arise from the noise matrix resulting of the multi-qubit measurement process. {Reconstructing the multi-qubit noise matrix in Eq.~\eqref{eq:many-qubit-bayes} ideally requires full detector tomography, but this quickly becomes infeasible as measurement requirements grow exponentially with qubit count.} Consequently, it is common practice to assume uncorrelated readout noise, allowing the multi-qubit noise matrix to be constructed as a tensor product of the single-qubit noise matrices as

\begin{equation}
\Lambda = \Lambda^{(q_1)} \Lambda^{(q_2)}... \Lambda^{(q_{N_q})}.
\label{eq:noise-tensor}
\end{equation}

While making the tomographic step easier, even the multi-qubit noise matrix of Eq. \eqref{eq:noise-tensor} does not reduce the computational cost of implementing Eq. \eqref{eq:many-qubit-bayes}
and the Bayesian measurement error mitigation remains resource intensive and cannot be employed as it is. 
For all these reasons, we have developed a heuristic strategy that enables the use of the scheme in the multi-qubit case, as depicted in Fig. \ref{fig:sketch} (d).
The first key component involves reducing the number of populations $\rho_{\mathbf i}$ included in the iterative procedure. We achieve this by running the mitigation algorithm only on the subspace of multi-qubit state probabilities where noisy outcomes have been measured. In other words, we exclude from the Bayesian cycle any states for which no strings are obtained in the experiment, limiting it to the strings $\bm{i}$ for which $\rho_{\bm{i}}^{\mathrm{noisy}} \neq 0$. Subspace reduction is a common strategy for measurement error mitigation techniques which suffer from the same exponential scaling problem, as discussed in \cite{Pokharel2024,nation2021}.

While parameter reduction is crucial to correct measurement errors in multi-qubit outcomes, it does not make handling the multi-variable probability distribution of Eq. \eqref{eq:many-qubit-bayes} much easier. Therefore, the next ingredient is to effectively reduce the number of variables to be used when handling the probability distribution which undergoes the Bayesian update. We achieve this by demoting all but two probabilities, $\rho_i$ and $\rho_j$, to constant estimates $R_k$, and redefining the conditional probability function for measuring a specific multi-qubit string as

\begin{equation}
\mathrm{P} (\bm s|\rho_{\bm i} ,\rho_{\bm j} ) = \Lambda_{\bm s \bm i} \rho_{\bm i}+\Lambda_{\bm s \bm j} \rho_{\bm j} +\sum_{\bm k \neq \bm i,\bm j}\Lambda_{\bm s \bm k} R_{\bm k}.
\label{Eq:cond-prob-bin}
\end{equation}

For a single measurement, Eq. \eqref{Eq:cond-prob-bin} is still the conditional probability of measuring the multi-qubit string $\mathbf{s}$ when the correct probabilities for strings $\mathbf{i}$, $\mathbf{j}$, and the remaining $\mathbf{k}$ (with $\mathbf{i} \neq \mathbf{j} \neq \mathbf{k}$) are $\rho_{\bm i}$, $\rho_{\bm j}$, and $R_{\bm k}$ respectively. However, only the first two are treated as variables of the probability distribution, while the remaining ones are kept constant. Using this form for the conditional probability, we can apply the Bayesian iterative cycle in the same fashion through

\begin{equation}
\begin{aligned}
\mathrm{P}(\rho_{\bm i} ,\rho_{\bm j} | \mathbf s^{(n)}) \propto  & \left [ \Lambda_{\bm s \bm i} \rho_{\bm i} + \Lambda_{\bm s \bm j} \rho_{\bm j} + \sum_{\bm k \neq \bm i, \bm j} \Lambda_{\bm s \bm k} R_{\bm k} \right ] \
\\& \times \mathrm{P}(\rho_i,\rho_j| \mathbf s^{(n-1)}),
\end{aligned}
\label{eq:many-qubit-bayes-simp}
\end{equation}

and obtain a final probability distribution for the two variables. The posterior probability distribution $\mathrm{P}(\rho_{\bm i} ,\rho_{\bm j} | \mathbf s^{(\mathrm N_{\mathrm {shots}})})$ can then be used to estimate the correct values for the probabilities of measuring  ${\bm i}$ and ${\bm j}$, thus obtaining some updated values for $R_{\bm i}$ and $R_{\bm j}$. {Furthermore, throughout our work, we sample the probability distributions for $\rho_{\bm i}$ and $\rho_{\bm j}$ on a grid with $n_p=R_{\bm i}+R_{\bm j} $ points}. After this step, the strategy is to repeat the procedure for a new pair of probabilities until all possible pair combinations have been updated once. {It is possible to further reduce the numerical complexity by considering pairs whose bit strings are within Hamming distance $d$, e.g. $H_d(i,j) \le d$.} This sequential heuristic binary optimization is then repeated until a chosen convergence criterion is achieved. A simple rule of thumb is to stop the algorithm repetitions if the change in the obtained populations does not change with respect to the previous iteration within a chosen accuracy. All the steps of the algorithm are outlined in Algorithm \ref{alg:Bayes}.

We have outlined the general theory behind Bayesian measurement error mitigation using a given bit string outcome $\bm s$. However, this framework can also be employed when we have access to analog detector data before any assignment to 0s and 1s is made. For example, in the case of superconducting qubits, the readout step exploits the dispersive interaction between the superconducting qubit and a readout resonator, which is the actual component of the QPU targeted by the readout pulse \cite{Blais2004,Blais2021}. Each measurement registers a point in the IQ-plane for each shot, which is then converted to a binary outcome, often according to its relative position with respect to some threshold value, as shown in Fig. \ref{fig:clouds}. Within the Bayesian mitigation framework, we can leverage the full continuous range of the physical measurement, albeit with the additional cost of calibrating the detector response, i.e. performing detector tomography, accordingly.

This means that if the detector measures the physical value $Q$, the detector tomography aims to reconstruct a noise matrix with a continuous index $\Lambda_{Qj} \equiv \Lambda_{j}(Q)$, which can be interpreted as the probability of measuring $Q$ if the qubit was in $| j \rangle$. The Bayesian measurement error mitigation algorithm can be applied in the same fashion as for bitstring measurements, noting that in the multi-qubit case we collect the set $\bm Q = Q^{(1)},...,Q^{(\mathrm N_q)}$ instead of the strings $\bm s = s^{(1)},...,s^{(\mathrm N_q)}$  for each shot. Thus, the Bayesian update rule makes use of the conditional probability
\begin{equation}
\mathrm{P} (\bm Q|\rho_i,\rho_j) = \Lambda_{\bm i} (\bm Q) \rho_{\bm i}+\Lambda_{\bm j} (\bm Q)\rho_{\bm j} +\sum_{\bm k \neq \bm i,\bm j}\Lambda_{\bm k}(\bm Q) R_{\bm k},
\label{eq:analog-cond-prob}
\end{equation}
which is the equivalent of Eq. \eqref{Eq:cond-prob-bin}.

In the following sections, we will demonstrate that the detector analog data, when available, contains more information than binary outcomes and its use enhances the effectiveness of the mitigation algorithm with negligible increase in computational time.

 \begin{algorithm}
\caption{Algorithm for multi-qubit Bayesian measurement error mitigation}
\label{alg:Bayes}
\begin{algorithmic}[1]
\REQUIRE Measurement outcomes for each shot, convergence tolerance $\epsilon$
\ENSURE Mitigated Multi-qubit state populations 
\STATE Compute initial noisy multi-qubit state populations $R_k \neq 0$, with $k=1,...,M \le 2^{\mathrm{N}_q}$
\STATE Reduce subspace to mitigate to $M$ states 
\WHILE {$\Delta (R^{(n+1)},R^{(n)}) \ge\epsilon$}
\FORALL{$i \neq j$, with $H_d(i,j) \le d$}
\STATE Promote $R_i$ and $R_j$ to variables $\rho_i$ and $\rho_j$
\STATE Perform Bayesian cycle to obtain $\mathrm{P}(\rho_i,\rho_j)$ 
\STATE Update $R_i$ and $R_j$ 
\ENDFOR
\STATE Update $M$ {\bf if} some $R_i$ becomes zero 
\STATE Compute $\Delta (R^{(n+1)},R^{(n)})$
\ENDWHILE
\end{algorithmic}
\end{algorithm}

Finally, it is worth mentioning that our approach to measurement error mitigation through Bayesian inference shares some similarities with the Iterative Bayesian Unfolding (IBU). IBU is a form of regularized matrix inversion applied to the noise matrix \cite{Nachman2020,Hicks2021}. The IBU mitigation scheme starts with the noise matrix $\Lambda$ and the noisy empirical distribution $\rho^{\mathrm{noisy}}$, and by applying Bayes’ rule to an initial guess $\rho_j^{0}$ it calculates the mitigated probability vectors $\rho_j^{n+1}$ via iteration, i.e.,

\begin{equation}
\rho_j^{n+1} = \sum \rho_i^{\mathrm {noisy}} \frac{\Lambda_{ij}\rho_j^{n}}{\sum_m \Lambda_{im}\rho_m^{n}}.
\label{eq:ibu}
\end{equation}
However, while the IBU method uses Bayes’ rule and starts with a prior guess, it does not result in a proper posterior distribution in the Bayesian sense.  For this reason, it has been more appropriately referred to as iterative expectation maximization unfolding, and it is worth mentioning that more recently the IBU update rule has been vectorized to enable fast parallel computation on GPUs \cite{Pokharel2024}.

\begin{figure}[!t]
\begin {center}
\includegraphics[width=0.4\columnwidth]{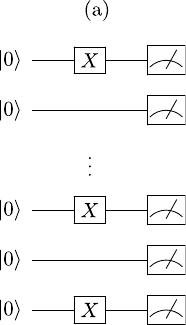}
\vspace{1cm}

\includegraphics[width=0.95\columnwidth]{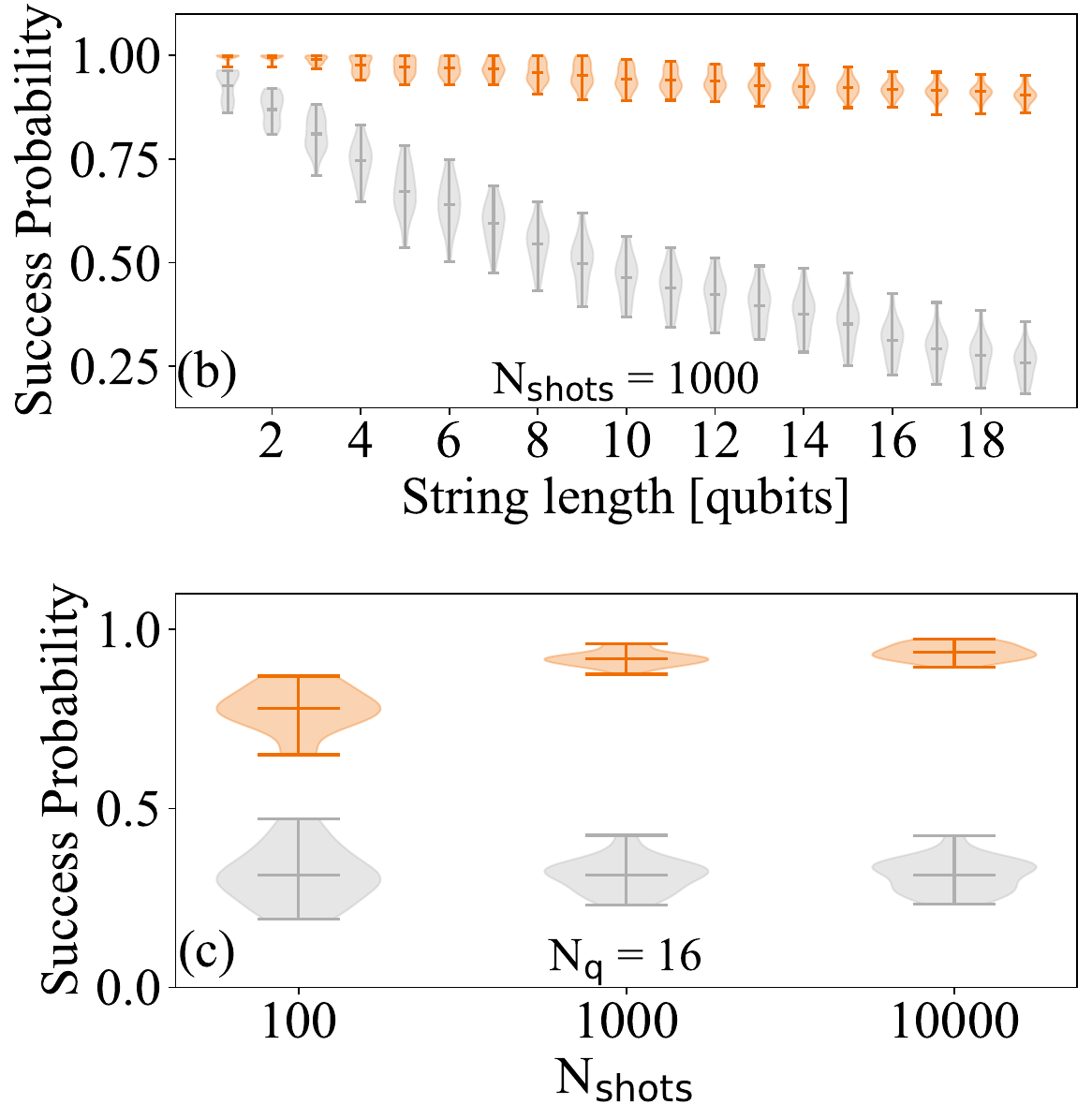}

\end{center}
\caption {(a): Quantum circuit for the preparation of a random multiqubit string: single qubit X gates are applied on the specific qubits ti flip. (b): Comparison between the mitigated (orange) and unmitigated (gray) success probability of measuring the prepared bitstring as a function of the string length. (c): Comparison between the mitigated (orange) and unmitigated (gray) success probability of measuring a prepared 16 qubits bitstring as a function of the number of shots. {The distribution shown in (b) and (c)  correspond to 20 different random bitstrings, with the middle bar representing the mean across the different realization, while the upper and lower bars are the highest and lowest success probabilities.}}
\label {fig:state_prep}
\end{figure}

\begin{figure*}[!t]
\begin {center}
\includegraphics[width=1.75\columnwidth]{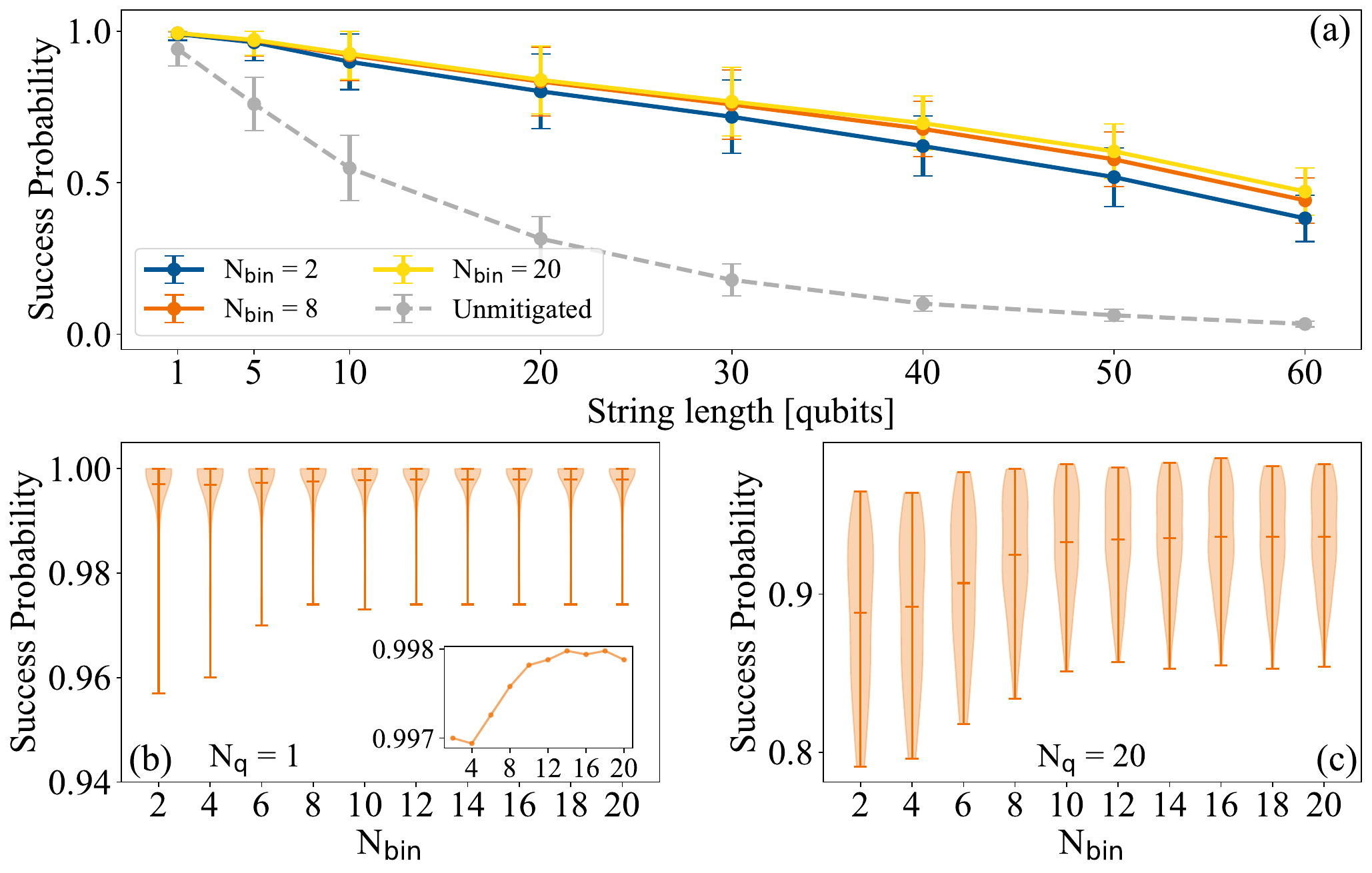}

\end{center}
\caption {{(a): Comparison of mitigated success probabilities for measuring a prepared bitstring as a function of string length, for different number of partitions of the Q-axis. The results are obtained with $10^3$ shots and averaged over 50 realizations, with error bars corresponding to the standard deviations.
(b): Success probability of measuring a single qubit as a function of $\mathrm{N_{\mathrm{bin}}}$ with the inset showing only the average values. (c): Success probability of measuring a 20-qubit string as a function of $\mathrm{N_{\mathrm{bin}}}$. The results shown in (b) and (c) are obtained with $10^3$ shots and show the results for all the 50 realizations. The middle horizontal bar represent the average mitigated success probability, while the upper and lower bars are the maximum and minimum of the measured samples.}}
\label {fig:analog}
\end{figure*}

\section{Bayesian mitigation with binary outcomes}
\label{scaling}

In this section, we apply our Bayesian measurement error mitigation strategy to binary multiqubit outcomes obtained by executing quantum circuits on the superconducting qubits quantum computer VTTQ20. The protocol begins with single-qubit measurement tomography to determine the uncorrelated error rates. While the service provider typically provides these error rates, we prefer extracting them explicitly.
We characterize the detector by performing two simple experiments: qubit reset followed by measurement and qubit flip followed by measurement. Using $\mathrm{N}_{\mathrm{shots}} = 10^5$ shots for each experiment, we obtain a two-by-two noise matrix $\Lambda^{(q)}$ for each qubit in the QPU. With these, we have access to the multiqubit noise matrix and we are now able to apply the Bayesian measurement error mitigation algorithm to the outcome of any quantum circuit.

To showcase the effectiveness of the measurement error mitigation scheme, we select a benchmark algorithm designed to minimize errors from faulty gates and decoherence. Specifically, we prepare a specific bit string through a single layer of single-qubit X-gates, i.e {$\mathrm U(\mathbf s) \ket {\mathbf 0} =\otimes_{i} X^{s_i}_i\ket {\mathbf 0} $}. As illustrated in Fig. \ref{fig:state_prep} (a), for a chosen bit string (e.g., "10..101"), we apply X-gates to specific qubits to flip their state from "0" to "1". {The benchmark figure of merit is the success probability, which is the probability of measuring the specific string $\mathbf{s}$ after applying  $\mathrm U(\mathbf s)$ to the initial state}. In these circuits, noisy readout is the primary source of errors, with potential imperfections in the rotation gate calibration accounted for within the noise matrix. 

Thus, we prepare 20 random bit strings of different length, ranging from $\mathrm{N}_q = 1$ to $\mathrm{N}_q = 19$. After executing the quantum circuits and measuring the results, we obtain the unmitigated initial estimates. We then apply the Bayesian mitigation procedure outlined in the previous section, as described by Eqs. \eqref{eq:many-qubit-bayes}-\eqref{eq:many-qubit-bayes-simp}.

In Fig. \ref{fig:state_prep} (b), we display the success probabilities obtained over 20 realizations as a function of the multiqubit string length, comparing them against the unmitigated ones. 
We observe that the unmitigated success probabilities decrease rapidly as the string length increases, highlighting the challenge posed by measurement errors in longer sequences. However, applying Bayesian measurement error mitigation results in significantly higher success probabilities across all lengths. In particular, for the longest string considered (19 qubits), the average success probability improves from approximately $26 \%$ without mitigation to $92 \%$ after mitigation with only $10^3$ shots. {Furthermore, we observe that mitigated values have a narrower distribution around the mean for each qubit number.}

In Fig. \ref{fig:state_prep} (c), we focus on a fixed multiqubit string length of \(\mathrm{N_q} = 16\) and plot the mitigated and unmitigated success probabilities as a function of the number of shots. We observe that, while the unmitigated success probability shows only slight improvement with an increasing number of shots, the mitigated probability experiences a substantial enhancement. For instance, with $10^4$ shots, the mitigated success probability reaches approximately $93\%$. This demonstrates the significant effectiveness of the Bayesian error mitigation technique in improving the accuracy of quantum measurements, even as the number of shots increases.

In the context of this string preparation circuit, the unmitigated result, on average, tends to plateau near the product of the individual readout fidelities, which for the 16 qubits used in this experiment corresponds to approximately $\sim 1/3$.
In contrast, the Bayesian measurement error mitigation algorithm continuously refines the mitigated outcome with each shot, offering further advantages in its application. Although the rate of improvement diminishes over time, this behavior is expected as Bayesian inference benefits from a larger number of measurements, allowing for a more precise estimation of the multiqubit state populations.

\section{Bayesian mitigation with Analog data}
\label{analog}
In this section, we demonstrate how readout analog data from the detector can boost the accuracy of the Bayesian measurement error mitigation scheme. Here, to validate these improvements, we run the algorithms on the superconducting qubits quantum computer "Nazca" - IBM127 (for details on calibration and QPU figures of merit see the Supplemental Material \cite{ibmcalibration}).

The initial step involves performing measurement tomography to reconstruct the detector response functions. To do this, similar to measuring error rates, we execute two algorithms: one where we measure the qubit after reset, and one where we measure the qubit after applying an X-gate flip.
In a superconducting qubit device, the typical measurement outcome is illustrated in Fig. \ref{fig:clouds} (top panel), where each measurement is depicted as a dot in the I-Q plane. Based on prior calibration, the standard method for assigning a measurement outcome as either “0” or “1” involves defining some threshold values or curves in this two-dimensional space. This assignment method is clearly not perfect. Instead of producing two distinct and sharp signals corresponding to the qubit's two possible states, the detector's response functions often exhibit significant overlap. This overlap is evident in the projected distributions shown in Fig. \ref{fig:clouds} (bottom panel), where the signals for the "0" and "1" states are not completely separated, causing then the assignment errors.

 When dealing with readout analog data, the reconstruction of these histograms is analogous to the process of reconstructing the noise matrix for binary outcomes. Formally, the histogram shown in Fig. \ref{fig:clouds} can be described by a function such as
\begin{equation}
\Lambda_j(Q) = \sum_{i=1}^{\mathrm{N_{\mathrm{bin}}}} \lambda_{ji} \mathbb{I}{[Q_i,Q_{i+1}]}(Q),
\end{equation}

where $\Lambda_j(Q)$ represents the probability of the state being $j$ when the detector measures the physical value $Q$, and $\mathbb{I}{[Q_i,Q_{i+1}]}(Q)$ is the indicator function defined as $\mathbb{I}{[Q_i,Q_{i+1}]}(Q) = 1$ if $Q \in [Q_i, Q_{i+1}]$ and $0$ otherwise. $\lambda_{ji}$ is then the probability of the state being $j$ when the measured $Q$ is in the i-th interval $\left [Q_i,Q_{i+1} \right]$. Here, $\mathrm{N_{\mathrm{bin}}}$ denotes the number of intervals used to discretize the Q axis of the IQ plane and serves as a parameter representing the discretization applied in the modeling of the detector response functions. Although an analytical expression for these distributions would be ideal, and although a bimodal Gaussian can often adequately capture the behavior of the readout functions, it is typically simpler to use directly the histograms. This approach is somewhat similar to defining Q-dependent error rates along the Q-axis partition.

Therefore, once we collect data from the $\mathrm{N_{\mathrm{shots}}} = 10^5$ shots, we establish a number of partitions, denoted as ${\mathrm{N_{\mathrm{bin}}}}$, and reconstruct response functions for each individual qubit. These response functions are then used in the multiqubit conditional probability of Eq. \eqref{eq:analog-cond-prob} to be used in the the update cycle of the Bayesian measurement error mitigation algorithm. In this framework, when we consider ${\mathrm{N_{\mathrm{bin}}}} = 2$, this approach is equivalent to working with binary outcomes.

\begin{figure*}[!t]
\begin {center}
\includegraphics[width=1.75\columnwidth]{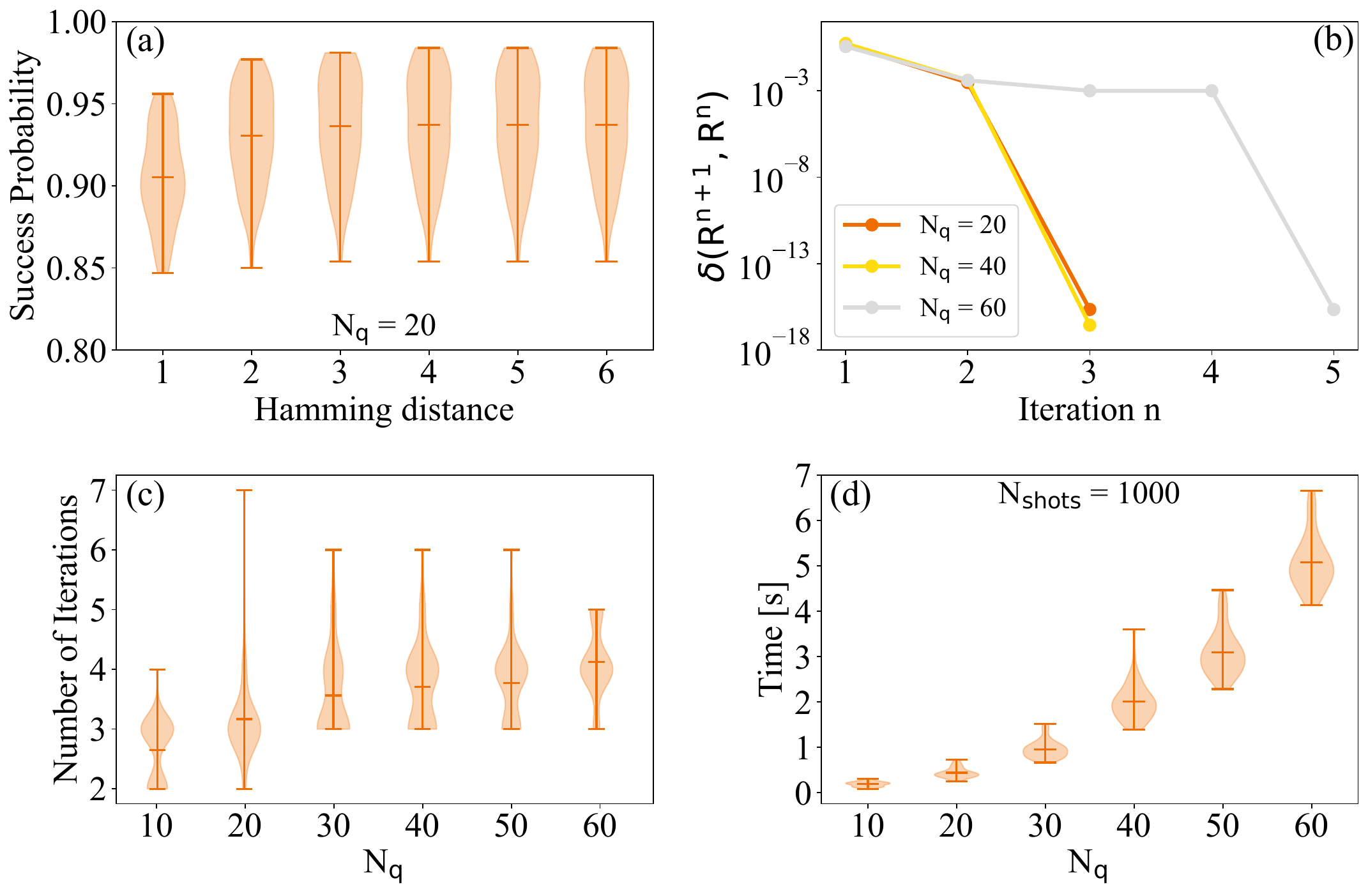}
\end{center}
\caption {{(a): Mitigated success probability as a function of the Hamming distance between population pairs undergoing the Bayesian update cycle. (b): Total variation distance between the mitigated probabilities after each Bayesian iteration as a function of the iteration number for different string lengths. (b): Number of iterations required by our Bayesian mitigation scheme to reach the convergence condition. (c): Time required by our Bayesian mitigation scheme to mitigate the outcomes of the string preparation experiment as a function of the number of qubits measured. The results in (a), (b), and (c) show the results obtained for 50 realizations with the middle horizontal bar line representing the mean of the distributions. In all cases the mitigation algorithm is applied using $\mathrm{N}_{\mathrm{bin}}=20$ as discretization parameter for the detector response function.}}
\label {fig:scaling}
\end{figure*}

In Fig. \ref{fig:analog}, we investigate the role of the number of partitions, ${\mathrm{N_{\mathrm{bin}}}}$, on Bayesian measurement error mitigation when using analog data. We prepare 50 random bit strings across various qubit lengths and record the measurement outcomes in analog form. After obtaining the unmitigated initial estimates, we apply the Bayesian measurement error mitigation procedure, leveraging the previously characterized Q-dependent detector response functions. Our aim is to understand how different values of ${\mathrm{N_{\mathrm{bin}}}}$, i.e. the granularity of the analog measurement data, can impact the mitigation process.
{In Fig. \ref{fig:analog} (a), we display the averaged mitigated success probabilities as a function of the string length for different values of ${\mathrm{N_{\mathrm{bin}}}}$ with the respective standard deviations. We see that increasing the granularity of the detector response function constantly result in better mitigated values for all  string lengths. For the case of $\mathrm N_q=20$, the success probability improves significantly from approximately $\sim 89\%$ to about $\sim 94\%$ when the discretization of the response functions increases from ${\mathrm{N_{\mathrm{bin}}}}=2$ to ${\mathrm{N_{\mathrm{bin}}}}=10$.
Both cases represent a substantial improvement over the unmitigated result which is around $\sim 25 \%$.
In Fig. \ref{fig:analog}(b), we show the mitigated single-qubit success probability for all the the 50 realisations as a function of ${\mathrm{N_{\mathrm{bin}}}}$. In the inset, which displays only the average values, we observe a noticeable trend where the success probability, which in the single qubit case is akin to the readout fidelity, improves as ${\mathrm{N_{\mathrm{bin}}}}$ increases.  In Fig. \ref{fig:analog}(c), we display the mitigated success probability for all the 50 realisations for a 20-qubit string. Also in this case, we observe that the mitigated success probability increases as the response functions are discretized with a greater number of partitions ${\mathrm{N_{\mathrm{bin}}}}$. However, these improvements eventually plateau beyond a certain threshold value of ${\mathrm{N_{\mathrm{bin}}}}$.}
These findings underscore how leveraging the additional information encoded in the analog data can enhance the effectiveness of the Bayesian measurement error mitigation algorithm with minimal to none additional computational cost. 
\subsection{Algorithm convergence and scaling}
\label{other}
In this subsection, we estimate the computational complexity and cost of the Bayesian multiqubit measurement error mitigation algorithm \ref{alg:Bayes} and applications of the previous subsection. 
A single Bayesian iteration, as described in Eq. \eqref{bayes-update}, scales linearly with the number of shots $\mathrm{N}_{\mathrm{shots}}$. This process must be repeated for each pair of states in the truncated subspace. If initially we have $M$ strings with counts such that $\rho_{\bm i}^{\mathrm {noisy}} \neq 0$, we have $\frac{M(M-1)}{2}$ populations pairs to consider and update. Thus, if all pairs are updated $N_r$ times, the overall computational complexity to post-process the results of a multiqubit experiment scales as $\propto \frac{N_r\mathrm{N}_{\mathrm{shots}} M (M - 1)}{2}$. 
This indicates a linear relationship with the number of shots and iterations and a quadratic relationship with the number of states after truncation based on the initial noisy counts. Therefore, the primary computational cost for large systems arises from the number of initial noisy populations, which is intricately linked to the number of shots and the extent to which the quantum state is spread over the multiqubit Hilbert space. Hence, the algorithm is more efficient when the outcome of the quantum algorithm is concentrated on a limited number of populations. {Thus, it is convenient to introduce another restriction and send through the Bayesian update cycle only pairs whose bit strings  are within a certain Hamming distance $d$, e.g. $H_d(i,j) \le d$. In Fig.  \ref{fig:scaling}  (a), we display the mitigated success probability for 20 qubits strings as a function of the Hamming distance used for selecting the qubit pairs. We see how, on average, there is negligible improvement after a Hamming distance of 3, which is then the threshold we set in all the executions of the Bayesian mitigation algorithm if not stated otherwise. }

Fig. \ref{fig:scaling} (b) shows the total variation distance between the probability vectors before and after each Bayesian update from Eq. \eqref{eq:many-qubit-bayes-simp} for each population pair. 
 The results shown here for a single realisation for qubit counts of $\mathrm N_q = 20, 40$
and 60 indicate that convergence is achieved after a few iterations as the figure of merit  {goes below $10^{-17}$}. Consequently, we set a threshold to terminate the Bayesian measurement error mitigation algorithm when the total variation distance between probability vectors before and after an iteration falls below $\epsilon = 10^{-8}$. To ensure the algorithm exits the iteration loop, we imposed a maximum of 20 iterations, which was never reached in our simulations.

{With this exit condition, we recorded the number of iteration required for our Bayesian mitigation algorithm to converge. In Fig. \ref{fig:scaling} (c), we show the number of iterations required for all 50 realizations and notice how the algorithm reaches the thresholds within 7 steps for all the case considered here. On average, it seems that the number of iterations slowly increases with the number of qubits.}

In Fig. \ref{fig:scaling} (c), we display the corresponding time required  to  mitigate the outcome of the random string state preparation circuit as a function of the number of qubits. The timing includes all the steps necessary to perform the mitigation algorithm as detailed in \ref{alg:Bayes}. Using a conventional quad-core laptop, the post-processing of the experiment data was on average within 5 second in the worst case considered here, i.e. $\mathrm N_q = 60$ and $\mathrm N_{\mathrm {shots}} = 1000$.

\begin{figure*}[!t]
\begin {center}
\includegraphics[width=1.75\columnwidth]{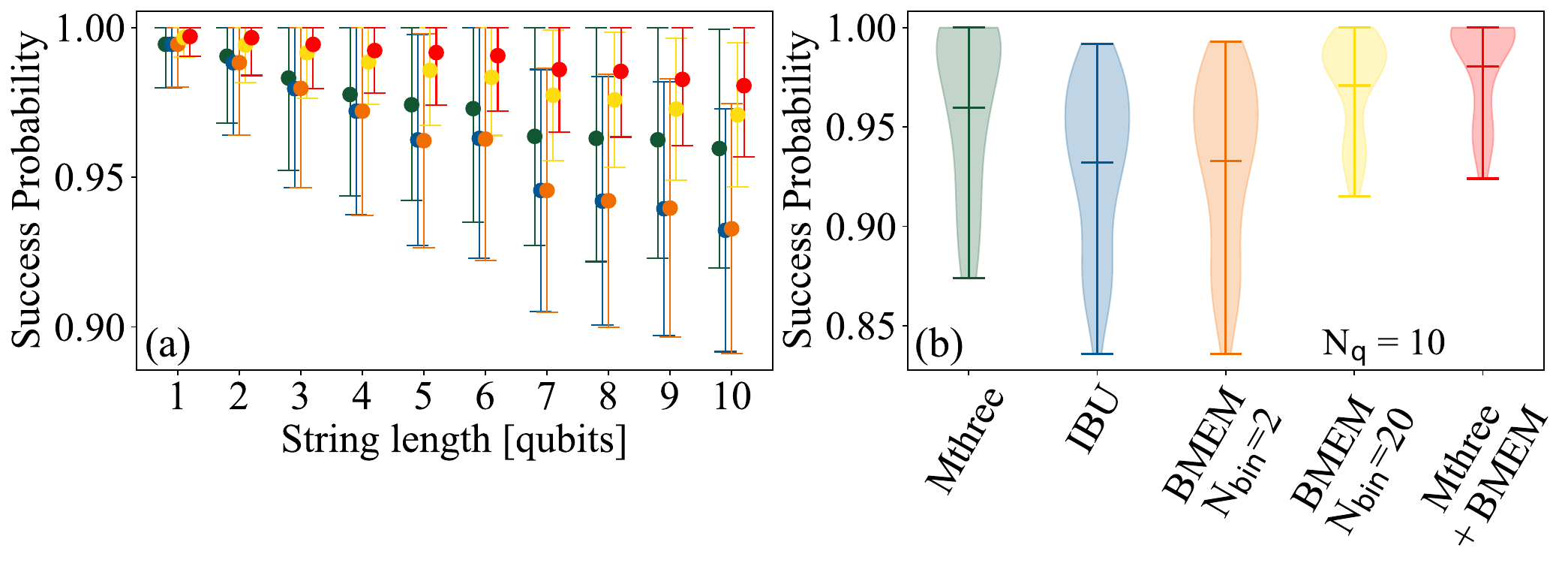}
\includegraphics[width=1.75\columnwidth]{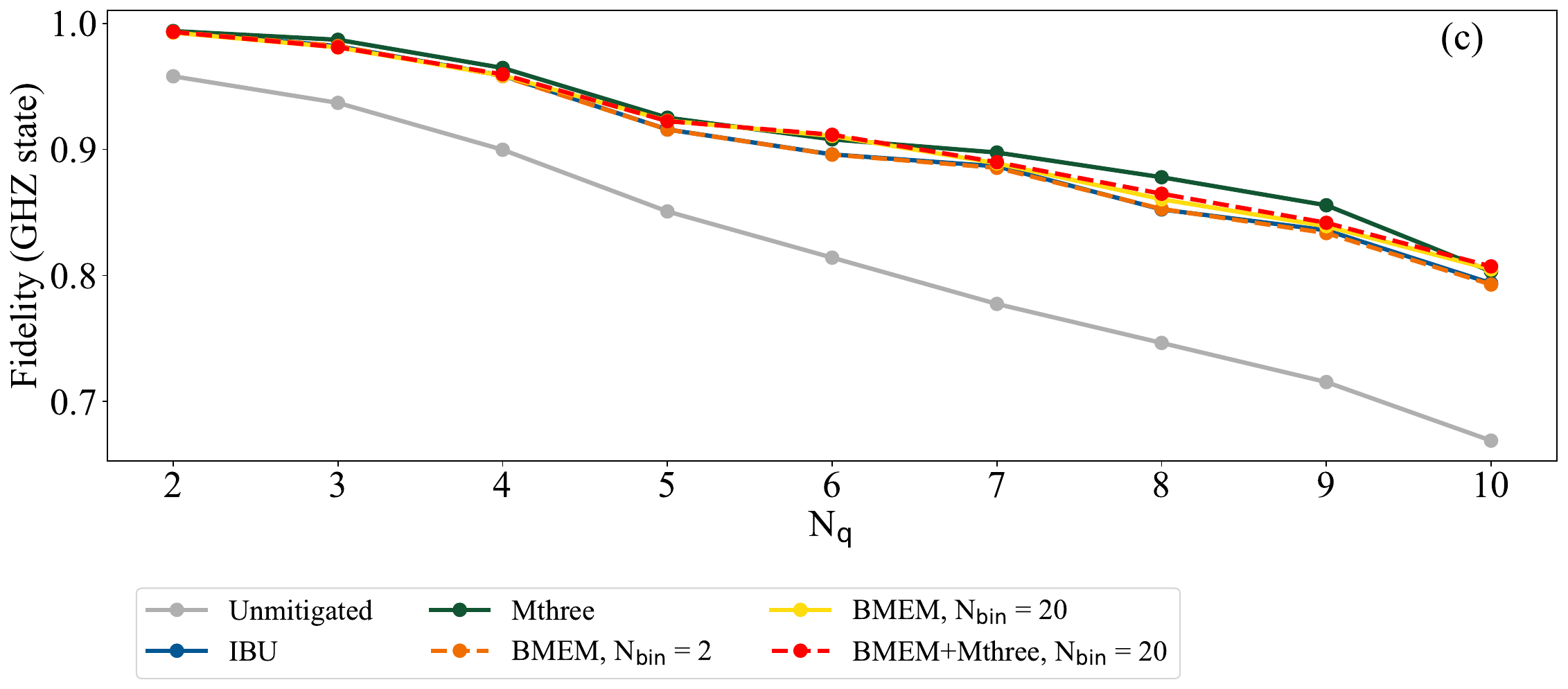}

\end{center}
\caption {{(a): Comparison between success probabilities of measuring the prepared bit-strings as a function of the string length mitigated according to different readout error mitigation strategies.  The results displayed are obtained by averaging over 50 realisations with error bars corresponding to the standard deviations. (b): Distribution over the 50 realisations for the mitigated success probabilities the different mitigation algorithms. (c) Comparison between the unmitigated and mitigated GHZ state fidelity calculated according through multiqubit coherence as a function of the number of qubit and mitigated through different mitigation strategies. The results shown in (b) and (c) are obtained with $10^3$ shots. The data shown in (a) and (b) was obtained on IBM127, while circuits for (c) were executed on VTTQ50.}}
\label {fig:ibu}
\end{figure*}

\section{Comparison and Integration with other mitigation techniques}
\label{comparison}

{The previous sections provide evidence about the performances of our implementation of Bayesian measurement error mitigation when applied to multiqubit experiments.} However, it is important to compare our method with other established measurement error mitigation techniques and for other quantum algorithms. For benchmark purposes, we selected two techniques. The first benchmark measurement error mitigation scheme is the Iterative Bayesian Unfolding algorithm (IBU) introduced at the end of Sec. \ref{theory}, which shares a probabilistic formulation with our protocol, linking correct qubit populations with the measured outcomes. 
However, IBU employs iterative expectation maximization rather than a formal prior-posterior update iteration in the Bayesian sense. {The second measurement error mitigation technique is the Mthree method \cite{nation2021}, based on noise matrix inversion. This method applies the inverse of the noise matrix to the measured outcomes in a reduced sub-space and, if the resulting mitigated probability vector is non-physical, finds the closest positive vector which still satisfies the sum-to-one condition.}
{
In Fig. \ref{fig:ibu} (a), we compare the success probabilities of measuring the prepared random bitstrings mitigated using IBU, Mthree, and our Bayesian measurement error mitigation method (dubbed in the figure as BMEM) as a function of the multiqubit string length, averaged over 50 realisations. All methods mitigate measurement errors efficiently, demonstrating their validity.  We apply three version of our Bayesian method: two cases where we apply our mitigation algorithm using the noisy counts as initial population pairs and two values for the discretization parameter $\mathrm N_{\mathrm {bin}}$ of the detector response function, i.e. 2 and 20; one case where the Bayesian mitigation algorithm uses the Mthree mitigated values as initial estimates for the multiqubit populations and $\mathrm N_{\mathrm {bin}}$=20. The workflow of the latter is depicted in Fig. \ref{fig:sketch} (c).}

\begin{figure}[!t]
\begin {center}
\includegraphics[width=0.95\columnwidth]{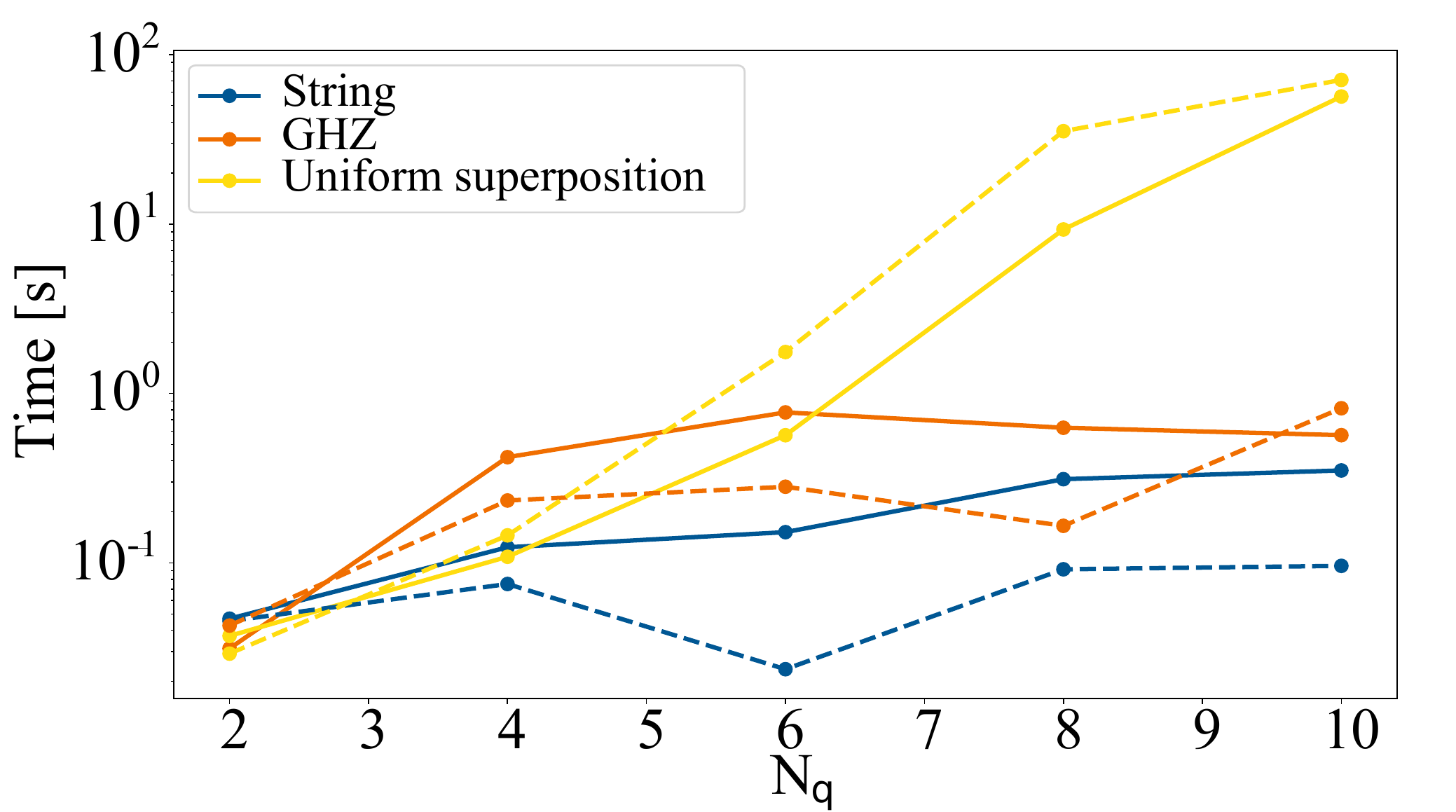}
\end{center}
\caption {{Time required by our Bayesian mitigation scheme to mitigate the outcomes for different quantum states as a function of the number of qubits. Solid lines consider the case where the Bayesian mitigation algorithm is applied starting from the noisy counts, the dashed to case where it is applied starting from the Mthree estimates as initial counts. All three state preparations were executed on VTTQ50 and $10^3$ shots.}}\label {fig:scaling_other_states}
\end{figure}

When applying our Bayesian mitigation algorithm to binary outcomes, i.e fixing $\mathrm{N_{bin}} = 2$, we achieve results very similar to IBU, with marginal improvements observed for the longer strings considered in our simulations. This similarity is expected because both methods start from similar probabilistic foundations, but employ different iterative approaches to reach the optimization/maximization.
On the other hand, by increasing the discretization parameter of the response function, our Bayesian method outperforms IBU. This enhancement suggests that our approach effectively leverages additional information encoded in the readout signal, which contributes to improved mitigated outcomes. Furthermore, Mthree seems to perform better than IBU, but on average worse than our Bayesian approach using analog data. Moreover, the highest accuracy is obtained by combining our Bayesian scheme using analog data with the Mthree mitigated outcomes as initial estimates for the Bayesian update cycle. In Fig. \ref{fig:ibu} (b), we display the distributions of the mitigated success probabilities for the $\mathrm{N_{q}} = 10$ qubits bitstrings where we observe how the different realisations are distributed around the average. It is worth to mention again how IBU and the Bayesian mitigation applied with binary outcomes have a very similar profile.
{As a next benchmark example, we prepare a GHZ state and calculate the state fidelity through the multiple quantum coherence \cite{Baum1985,Wei2020}. In Fig. \ref{fig:ibu} (c), we compare the unmitigated state fidelity with the ones obtained mitigating the circuits outcomes with the different mitigation techniques. All the techniques used show an improvement over the unmitigated result. As in the previous example, Mthree and Bayesian measurement mitigation with analog data seem to perform better than IBU, while Mthree results in the highest accuracy for most qubit numbers. 
To conclude our benchmark, in Fig. \ref{fig:scaling_other_states}, we compare the time employed by the Bayesian measurement error mitigation algorithm when mitigating the outcome of three different quantum states as a function of the qubit number. The three states considered are: random bit-string, GHZ state, and uniform superposition, i.e. the state obtained by applying an Hadamard gate to each qubit. We observe how the computational time is lower when mitigating single multiqubit strings, and increases when mitigating the outcome of a GHZ state and uniform superposition. The latter can be considered an upper bound, or worst case scenario, to the computational cost of the algorithm, as the solution is spread over the entire basis state, and the number of different basis states included in the mitigation algorithm is determined by the number of shots.}

\section{Conclusions}
\label{end}
In this work, we have introduced and conducted a comprehensive analysis of an efficient measurement error mitigation scheme that significantly enhances the accuracy of the readout step in multiqubit experiments on quantum computers. The algorithm combines Bayesian inference with a heuristic binary optimization cycle to construct a probability distribution for the multiqubit state populations, based on the known noise matrix affecting the measurement and state assignment. 
Crucially, our Bayesian framework is designed to avoid assigning nonphysical values, ensuring that the mitigated outcomes remain consistent with the experimental data at each step of the mitigation algorithm. We tested the scalability and performance of the algorithm on three real superconducting qubit quantum computers, demonstrating a substantial improvement in the precision of multiqubit bitstrings at the cost of a classical computational overhead.
Additionally, we have shown that it is more advantageous to apply readout mitigation techniques at the level of the detector's analog data. For superconducting QPUs, this involves working with data in the IQ-plane rather than aggregated binary counts. We compared our method against the state-of-the-art Iterative Bayesian Unfolding (IBU) mitigation technique and the Mthree method and found that our approach can outperform IBU, when incorporating analog readout data into the mitigation scheme. {Furthermore, we have shown how our method can be integrated with existing techniques to obtain even better performances.}
Although we focused on superconducting qubit devices, our mitigation approach is versatile and could be applied to other quantum computing platforms, either directly applying the mitigation procedure on the aggregated bitstring counts or using any detector analog data if available. This flexibility underscores the broad applicability and potential impact of our Bayesian measurement error mitigation framework in advancing the field of quantum computing.

\begin{acknowledgments}%
We acknowledge the use of IBM Quantum services for this work. The views expressed are those of
the authors, and do not reflect the official policy or position
of IBM or the IBM Quantum team. This work has been
partly funded by the Business Finland Co-Innovation Project
40561/31/2020 Quantum Technologies Industrial (QuTI). This research was partially supported by the PNRR MUR project PE0000023-NQSTI through the secondary projects ``ThAnQ" and ``QuCADD''.
\end{acknowledgments}

\bibliography{biblio}


\end{document}